\documentclass[twocolumn,dvipsnames]{aastex62}

\usepackage{CJK}
\usepackage{framed}

\graphicspath{{./}{figures/}}

\submitjournal{ApJ}

\def\Msun{{\rm\,M}_\odot}

\shorttitle{}
\shortauthors{Noel, Zhu and Gnedin}

\begin{document}
\begin{CJK*}{UTF8}{gkai}

\title{Mass-Metallicity Relation during the Epoch of Reionization in the CROC
Simulations}

\correspondingauthor{Isaac Noel}
\email{isaac18b@gmail.com}

\author{Isaac Noel}
\affiliation{New Trier High School; Winnetka, IL 60093, USA }

\author[0000-0003-0861-0922]{Hanjue Zhu (朱涵珏)}
\affiliation{Department of Astronomy \& Astrophysics; 
The University of Chicago; 
Chicago, IL 60637, USA}

\author{Nickolay Y.\ Gnedin}
\affiliation{Particle Astrophysics Center; 
Fermi National Accelerator Laboratory;
Batavia, IL 60510, USA}
\affiliation{Kavli Institute for Cosmological Physics;
The University of Chicago;
Chicago, IL 60637, USA}
\affiliation{Department of Astronomy \& Astrophysics; 
The University of Chicago; 
Chicago, IL 60637, USA}

\begin{abstract}
The low-redshift mass-metallicity relation (MZR) is well studied, but the high-redshift MZR remains difficult to observe. To study the early MZR further, we analyze the Cosmic Reionization on Computers (CROC) simulations with a focus on the MZR from redshifts 5 to 10. We find that, across all redshifts, CROC galaxies exhibit similar stellar-phase and gas-phase MZRs that flatten with higher stellar mass. We attribute this flattening to the inaccurate star formation and feedback modeling in CROC (star formation is overly suppressed in massive CROC galaxies). In addition, we show that the ratio between stellar metallicity and gas metallicity ($Z_*/Z_{gas}$) decreases 
as stellar age increases, meaning that in CROC galaxies, gas accretion rate is lower than metal production rate. With JWST we will be able to compare our predictions to observations of the Epoch of Reionization and understand better early galaxy formation.

\end{abstract}

\section{Introduction}\label{sec:intro}

As star formation continues in galaxies, stellar death through supernovae sends newly formed metals into the surrounding gas. The metals in the gas in turn enhance the metallicity of the next generation of stars. The metallicity of stars in a galaxy gives us information on when the galaxy acquired its star-forming gas, where stars are formed, and the impact of inflows and outflows on the mass and metals of the galaxy. 

A well-studied and well-observed property of galaxies at low redshift is the galaxy mass-metallicity relation (MZR). As discovered in both observations and numerical simulations, there is a tight correlation between galaxy mass and stellar metallicity as well as galaxy mass and gas metallicity. While we see an abundance of evidence for the MZR at low redshifts (\cite{Maiolino2} summarizes recent observational efforts, including \cite{Harikane, Faisst, Cullen}), it is uncertain whether this scaling relation exists at high redshifts, because strong emission lines used to measure galaxy metallicity at low redshifts, such as [OIII], are redshifted out of the current observational wavelength limit. However, with the successful commisioning of JWST \citep{Gardner2006,Rigby2022}, we will soon be able to study the high redshift MZR in detail.

To better interpret the forthcoming JWST data, we need theoretical understanding of the high-z galaxy properties and their relations. This is why we turn to numerical simulations to study the high redshift MZR. MZR at both low-z and high-z have been analyzed by multiple groups using simulations employing different numerical methods \citep[e.g.][]{Fire, Illustris, Langan, Henriques13, Henriques15, Lu}. Many of these works rely on ``zoom-in'' simulations, which treat galaxy feedback and baryon cycle models with great detail and produce realistic galaxies. However, they are limited to a small sample size of galaxies which may fail to demonstrate large-scale emerging patterns.

In this paper, we use the Cosmic Reionization on Computers (CROC) simulations \citep{gnedin14}, which we will explore further in the methodology section. The CROC simulations give a large sample of galaxies at $z>5$, and model well the relevant reionization physics. Therefore, CROC is suitable for studying the MZR during the Epoch of Reionization.

This paper has defined goals. We focus on the high redshift gas-phase and stellar-phase mass-metallicity relation and then make comparisons with current observations of nearby dwarf galaxies. We organize the paper as follows. Section~\ref{sec:methods} describes the CROC simulations, and how we measure the gas-phase and stellar-phase metallicities. In Section~\ref{sec:results}, we show both the gas-phase and stellar-phase MZR of the CROC galaxies. In addition, we study if the history of cosmic reionization affects the MZR.

\section{Methodology}\label{sec:methods}

\subsection{CROC Simulations}

We utilize data from the CROC simulations which use Adaptive Refinement Tree (ART) code \citep{ART1}. The CROC simulations are large-scale cosmological simulations that model the intergalactic medium (IGM) and resolve galaxies at high redshifts with a focus on cosmic reionization. They account for a plethora of physical processes relevant to cosmic reionization, including gas dynamics and star formation. More details are described in the first CROC methodology paper \citep{gnedin14}.

From CROC, we use six simulation boxes with volumes of $40 \,h^{-1} {\rm Mpc}$ comoving (cMpc). Each box includes simulation snapshots from redshifts 5 to 10. In the results we show, we include galaxies in all of the six boxes. We also analyze a seventh box that is strongly overdense so that we can analyze the impacts of reionization histories on the MZR. At $z = 5$, each box includes $1.4 *  10^5$ halos above the mass of $10^5 \Msun$. We use the Rockstar \citep{rockstar} halo finder to identify dark matter halos in CROC, and use the ``yt'' package to analyze these dark matter halos \citep{yt}.
To avoid errors that arise when the size of our halos approach the simulation resolution, we ignore halos with $M_{*} < 10^5 \Msun$.

\subsection{Measuring Stellar and Gas-Phase Metallicities}

We analyze the gas cells within a certain distance away from each dark matter halo center to determine the gas-phase metallicities ($Z_{\textrm{gas}}$). Gas-phase metallicity is mass-weighted. Since gas metallicity is related to star formation activity, we use only star particles within $R_{90}$, the radius within which contains 90\% of the stars in each galaxy. We also test our choice of radius by using only star particles within $R_{50}$ and see a negligible difference in the results.

Stellar metallicities ($Z_*$) are much simpler to obtain - we use all stellar particles within $R_{90}$  and measure $Z_*$ of a galaxy as a mass-weighted average of the stellar particles' metallicity.

Note that the metallicities we present in this paper are all in absolute units, not solar units.

We define the stellar mass of a halo, $M_*$, as the sum of stellar mass of all the star particles in the halo within the $R_{90}$ of the halo.

\section{Results}\label{sec:results}

In this section, we explore the gas-phase and stellar-phase metallicities of halos in the CROC simulation at high redshifts (5 $< z <$ 10) and examine how they are impacted by their intergalactic environment. We then show how $Z_\textrm{gas}$ and $Z_*$ evolve with redshift and DM halo age before comparing our results to observations.

\subsection{The Gas-Phase MZR}

\begin{figure}[htb!]
    \centering
        \includegraphics[width=0.99\columnwidth]{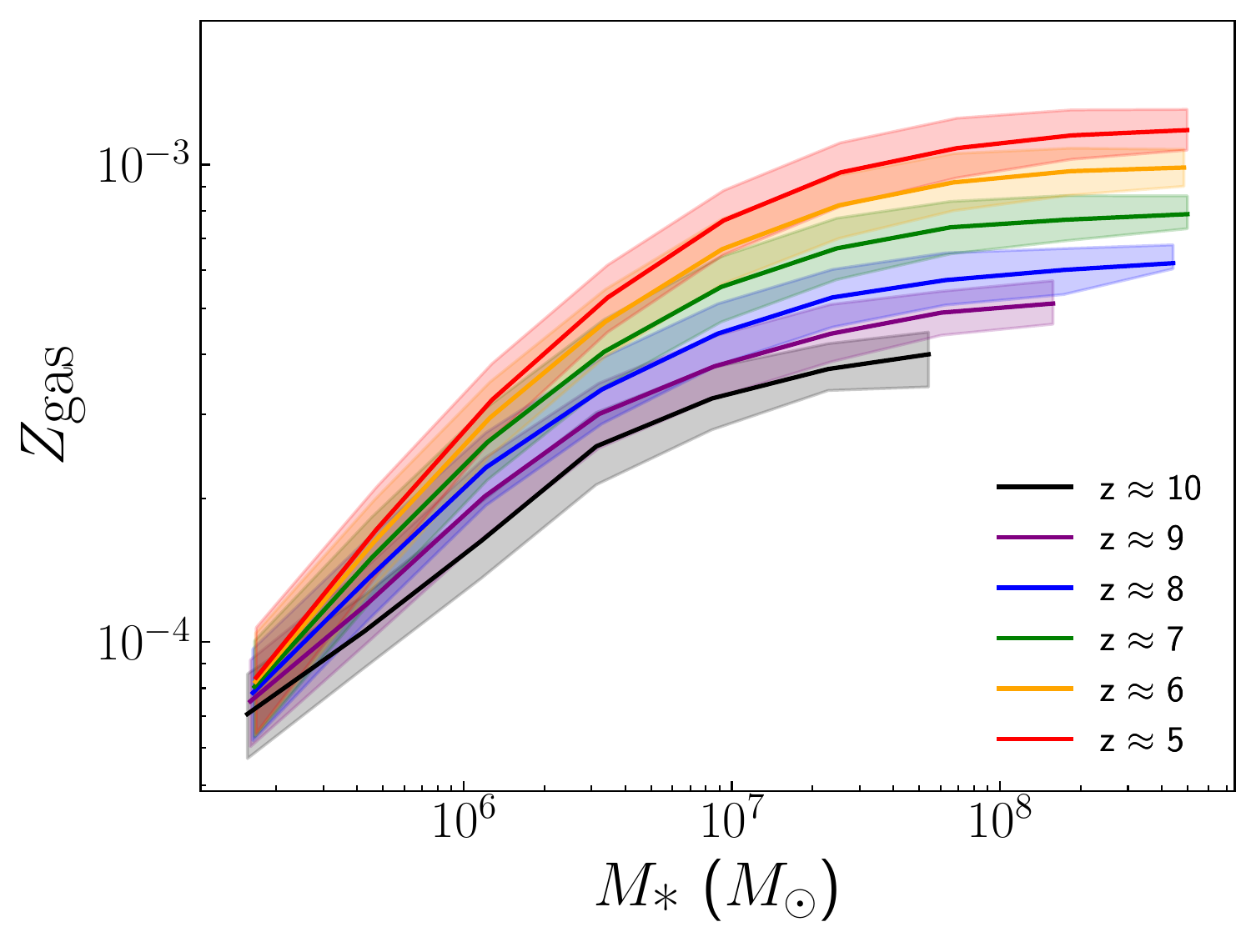} 
\caption{The evolution of the gas-phase mass-metallicity relation at different redshifts. We distinguish between redshifts by displaying lines of different color. The shaded regions represent the 25th to 75th percentile of the distribution. We find a clear MZR at redshifts 5 to 10, and that higher redshift galaxies have lower metallicities across our galaxy mass range.}\label{fig:mstar_zgas}
\end{figure}

In Figure~\ref{fig:mstar_zgas}, we see that the halos display a clear MZR which flattens with higher mass. This change in slope contrasts findings from the FirstLight Simulations \citep{Langan} that predict a flat MZR. In addition, we see that the MZR has a redshift dependence, with higher redshifts having lower metallicities. This trend agrees with predictions from IllustrisTNG and FIRE simulations \citep{Illustris, Fire}, and contrasts findings in \citet{Langan}.

Galactic chemical evolution is tightly connected to the interplay between star formation, gas outflows, and gas recycling and accretion. For example, the stellar mass-halo mass relation indicates that star formation efficiency increases with galaxy mass, suggesting that more massive galaxies should more metal-enriched. Meanwhile, galactic winds tend to blow away more gas and metals from lower mass galaxies because of their shallow potential wells. Our finding that gas metallicity increases with halo stellar mass is consistent with these models. The flattening of the MZR at high mass could be attributed to halo mergers; however, we note that in CROC, star formation is suppressed in massive galaxies, likely because the CROC model for star formation and feedback mis-models star formation in high-mass halos \citep{Zhu2020}.

The MZR serves as a test for star formation and feedback models implemented in the simulations. Comparison between current simulation predictions and future high-z observational data will inform us on the usefulness of our current theoretical models.

\subsection{The Stellar-Phase MZR}

\begin{figure}[htb!] 
    \centering
        \includegraphics[width=0.99\columnwidth]{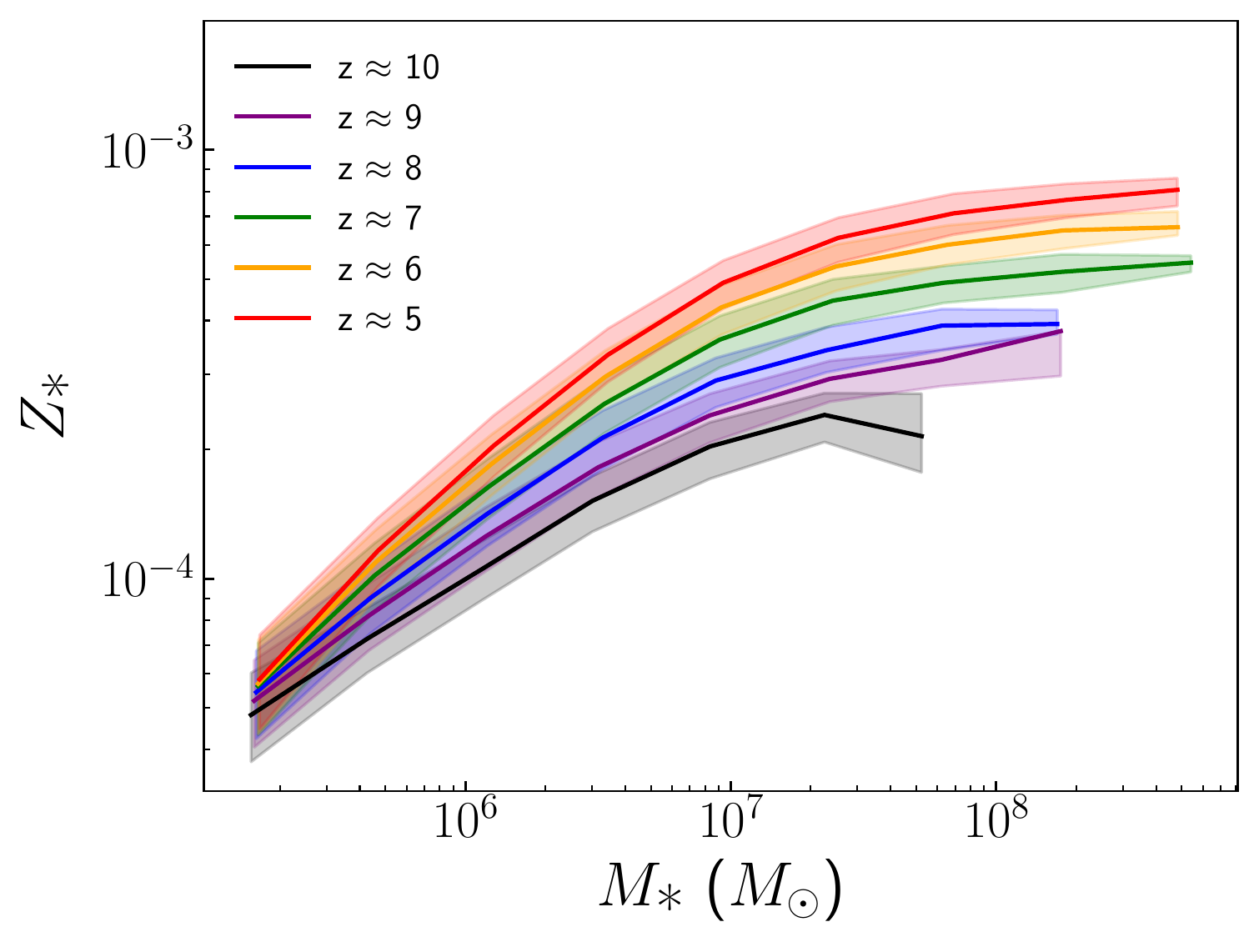} 
\caption{The evolution of the stellar-phase mass-metallicity relation at different redshifts. We distinguish between redshifts by displaying lines of different color. The shaded regions represent the 25th to 75th percentile of the distribution. We observe a similar trend and redshift dependence as the gas-phase MZR.  }\label{fig:mstar_zstar}
\end{figure}

We see a relation that is similar to the gas-phase MZR in Figure~\ref{fig:mstar_zstar}. The main difference is that the gas-phase MZR is higher than the stellar MZR at all redshifts. This difference is expected, as stellar metallicities reflect the time-averaged galactic metallicity across the galaxy star formation history, and gas metallicities give information about the current state of metal enrichment in galaxies. Combining this with the fact that star formation enriches the gas with metals as time progresses, we expect the stellar metallicity to be lower than the gas metallicity.

\subsection{Comparing $Z_*$ and $Z_\textrm{gas}$ as the DM halos age}

\begin{figure}[htb!]
    \centering
        \includegraphics[width=0.99\columnwidth]{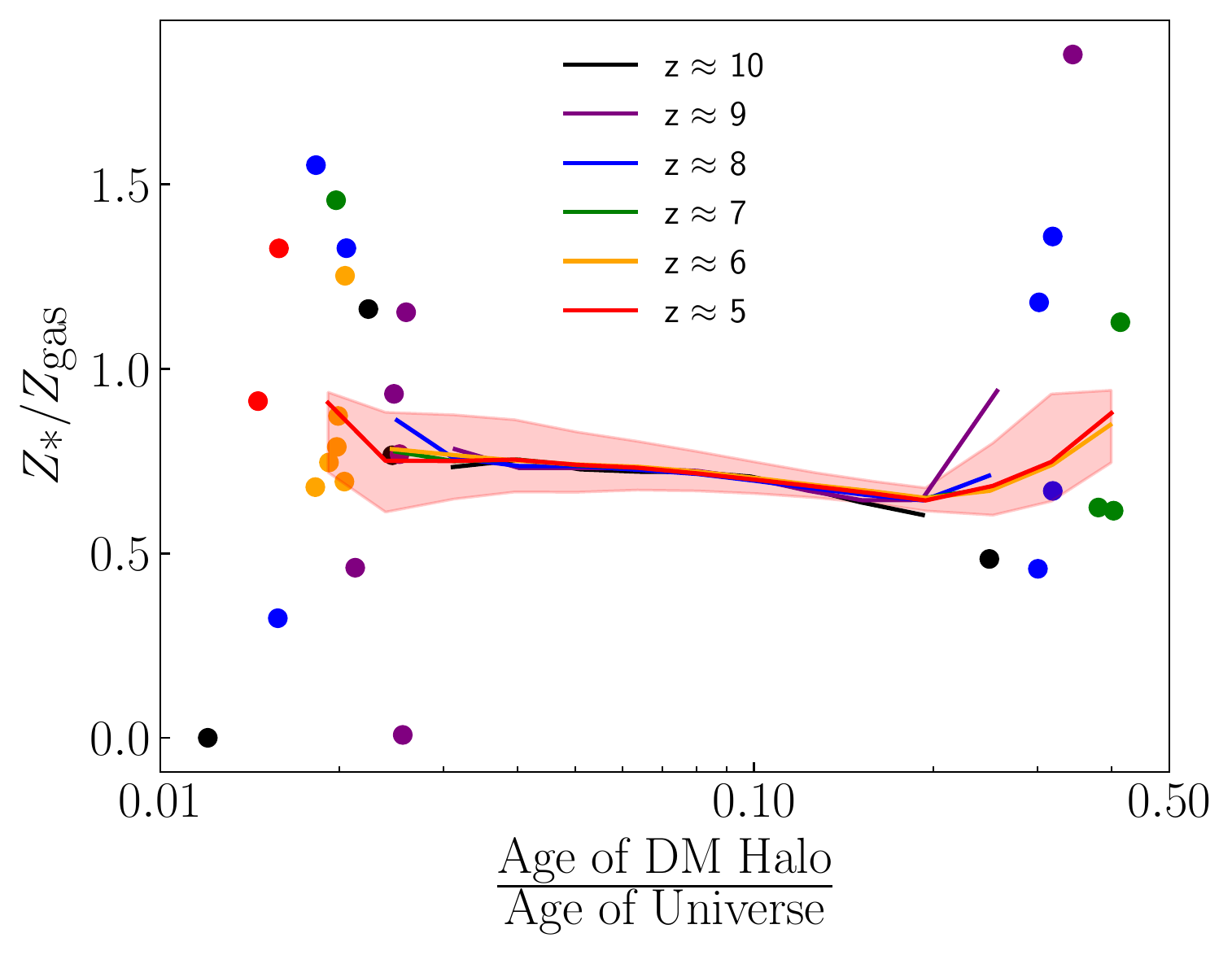} 
\caption{$Z_*/Z_{gas}$ as a function of normalized stellar age (age of halo / age of universe at redshift). The shaded region represent the 25th to 75th percentile of the distribution. The spread is similar across redshifts so we only show it at $z=5$. In bins with small sample sizes, individual galaxies are shown in points with colors corresponding to redshifts. Other than the sharp upward curve at high stellar ages (which can be attributed to outliers), the plot displays a downward slope.}\label{fig:age_zz}
\end{figure}

We now analyze the ratio between stellar metallicity and gas metallicity as halos age. This ratio gives information about the relative rate at which galaxies produce metals and replenish gas. We normalize the age of halos by the age of the universe at a given redshift. By normalizing the halo age, we can remove the redshift dependence that is inherent in the age of the universe. The age of the universe is calculated assuming a flat universe with $H_0$ = 68.14 km/s/Mpc and $\Omega_m = 0.3036$. As shown in Figure~\ref{fig:age_zz}, the ratio between $Z_*$ and $Z_\textrm{gas}$ decreases with the normalized halo age.

We see in the plot that $Z_*/Z_{gas}$ is independent of redshift. This means that the relative metal production and gas accretion rate is similar at $5<z<10$. The negative slope in the plot shows us that $Z_{gas}$ rises faster than $Z_*$. If metals entered the gas at a constant rate, the $Z_*/Z_{gas}$ ratio would be a constant because the gas metallicity and stellar metallicity would rise at the same rate. The negative slope tells us that as halos age, the metal production rate is higher than the rate at which the stars accrete pristine matter.

\subsection{Environmental effects on the MZR}

\begin{figure}[htb!]
    \centering
        \includegraphics[width=0.99\columnwidth]{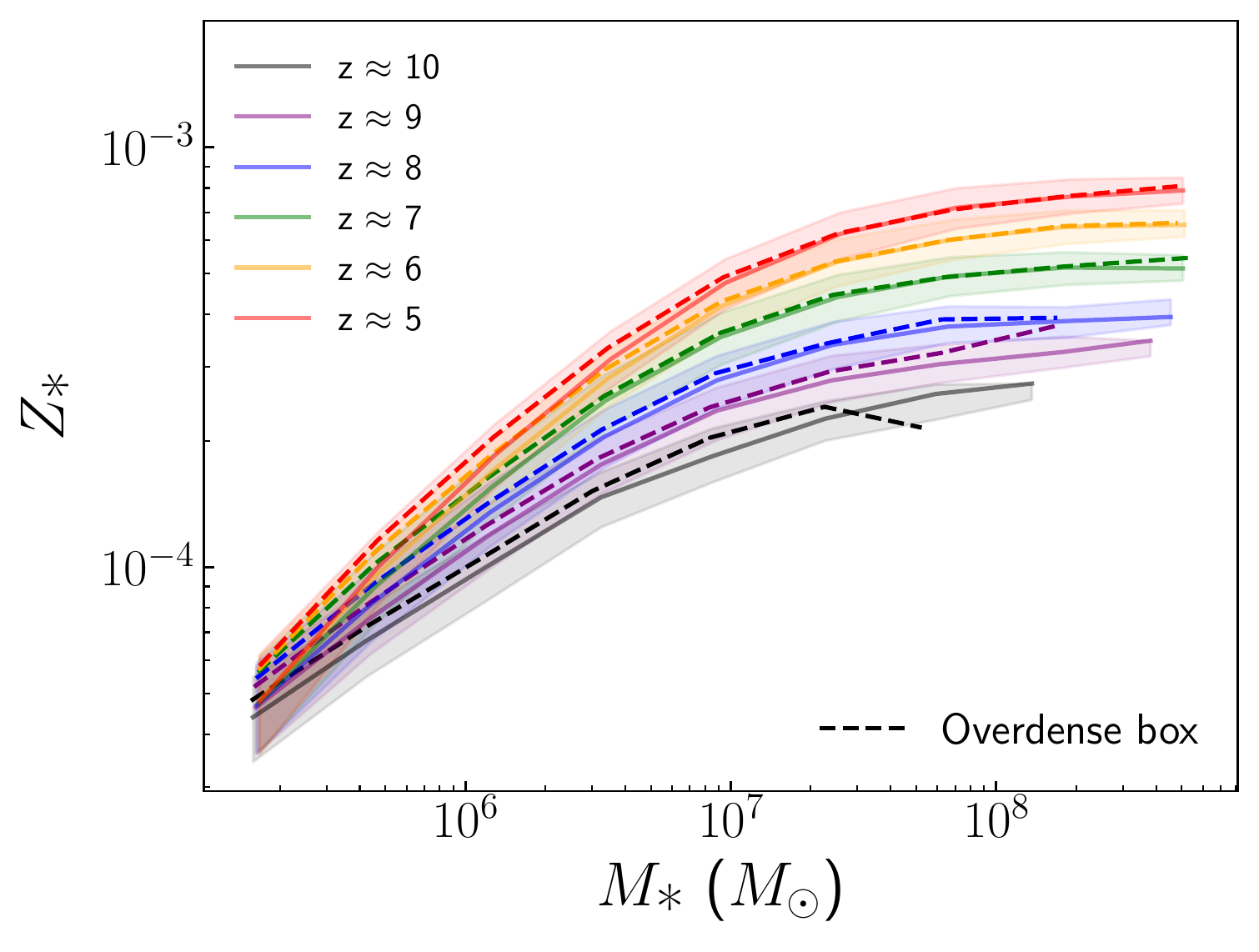} 
\caption{Stellar metallicity as a function of stellar mass. The dotted lines show the MZR in realistically dense simulations while the dotted lines show the MZR in overdense simulations.}\label{fig:mzs_edc}
\end{figure}

CROC simulations have high-fidelity radiative transfer modules implemented, making them suitable for exploring the effect of the global radiative feedback on the mass-metallicity relation.

To study how the intergalactic environment affects the MZR, we study DM halos from an overdense simulation. As shown in Figure~\ref{fig:mzs_edc} in dashed lines, the MZR in the overdense boxes is not very much different from the MZR observed in the realistic boxes. We see the same redshift dependence as well as the same flattening. From our results, we can conclude that the MZR is not impacted by intergalactic interactions and that it is likely driven by physical processes on a galactic scale, which agrees with the conclusions we have made so far.

\subsection{Comparison to Observation}

\begin{figure*}[htb!]
    \centering
    \begin{minipage}{0.49\textwidth}
        \centering
        \includegraphics[width=0.99\textwidth]{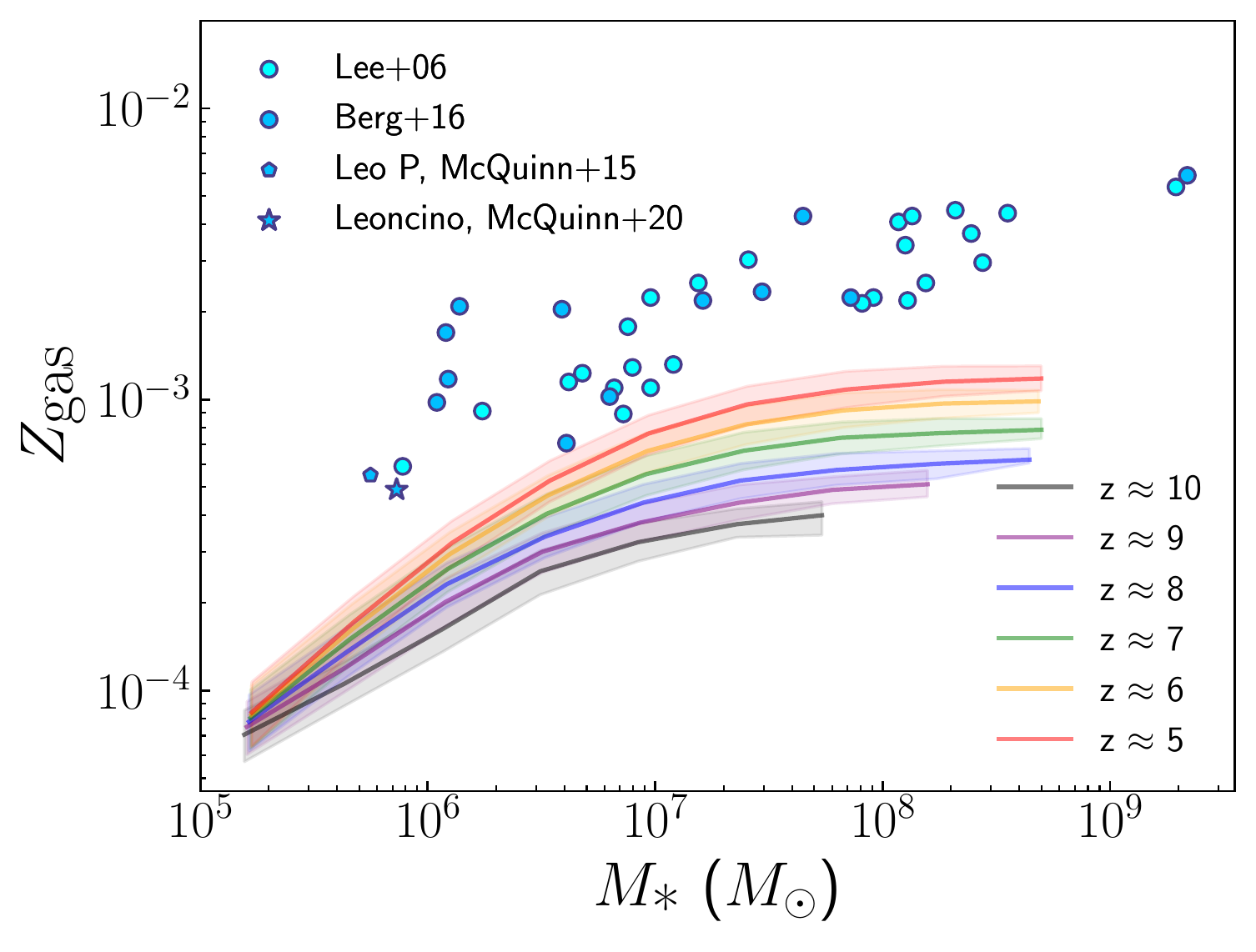} 
    \end{minipage}\hfill
    \begin{minipage}{0.49\textwidth}
        \centering
        \includegraphics[width=0.99\textwidth]{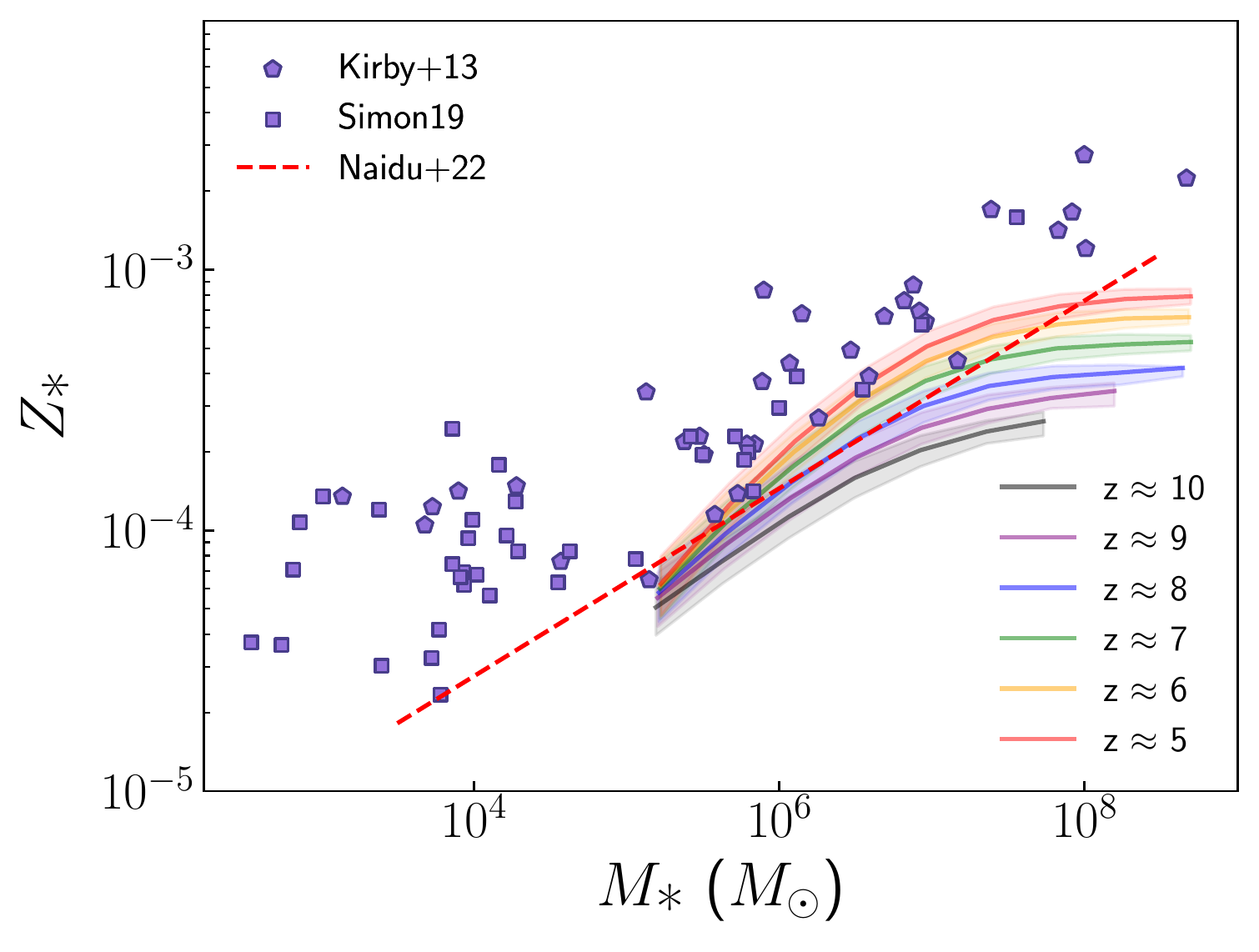}
\end{minipage}
\caption{Left: The gas-phase MZR from our simulations compared to the MZR observed in nearby dwarf galaxies. Right: The stellar-phase MZR from both our simulations and observations. The observational data points are from \cite{Lee, Kirby, McQuinn2015b, Berg, Simon, McQuinn2020}.}\label{fig:obs}
\end{figure*}

In this section, we compare our predictions with the MZR measured in nearby dwarf galaxies. Dwarf galaxies are considered to be the driving force of cosmic reionization and they are also most sensitive to the effects of reionization as the reionization process kills star formation in them. In Figure~\ref{fig:obs}, we plot our predictions along with observed gas and stellar metallicites from \cite{Lee, Kirby, McQuinn2015b, Berg, Simon, McQuinn2020}.

The left panel of Figure~\ref{fig:obs} shows the gas metallicity as function of stellar mass. Our predictions are reasonable when considering the observed MZR at low redshifts. The redshift dependence exhibited by the high-redshift galaxies continues into the present as the low-redshift MZR lies above the high-redshift MZR with reasonable distance. We also see the low-redshift gas metallicities take a similar shape to our predictions, exhibiting flattening at higher masses while maintaining a steeper rise around $M \approx 10^6 M_{\odot}$. However, when comparing to observation it's important that we acknowledge the systematic uncertainties present in both observations and simulations.

In the right panel of Figure~\ref{fig:obs}, we plot the stellar metallicity with stellar mass. Similarly to the gas-phase MZR, here is a fair agreement between observations and the simulation. However, we note that observationally, there is a much larger scatter in the dwarf galaxy stellar metallicities. A recent paper from \cite{Naidu_2022} has further made the distinction and showed that the tidally shredded ``disrupted dwarfs'' have lower metallicities by roughly 0.3 to 0.4 dex than the largely intact ``surviving dwarfs''.

\section{Summary and Discussion}

We have analyzed data from the CROC simulations to predict the mass-metallicity relation at high redshifts ($5<z<10$).

Our main results are:

\begin{enumerate}
    \item{The MZR exists at high redshifts (5-10).}
    \item{The MZR is not affected by the cosmic reionization histories.}
    \item{As galaxies age, they accrete metal-poor gas slower than they replenish the metals through star formation.}
    
\end{enumerate}

Now that JWST has successfully established itself at L2, we can expect high-quality spectroscopic data that allows high-z metallicity measurements. These observations will open the door to theories surrounding galaxy formation and evolution at cosmic dawn. 

From new empirical data, we can learn about how we predict the MZR.
In the literature, there are many predictions for the high-redshift MZR with all of them being from modern simulations. 
Each simulation uses different methods and prioritizes different physical processes. 
Confirming or denying any simulation's prediction will not only give us insight into the state of the universe at cosmic dawn, but it would also give us information that would lead to changes in our simulation methods. 
Perhaps we will see another MZR entirely, indicating that there is an aspect of high-redshift galaxies that we have yet to consider. 
There is much to discover about the universe, especially with the trove of JWST's new data.

\acknowledgments

Fermilab is operated by Fermi Research Alliance, LLC, under Contract No. DE-AC02-07CH11359 with the United States Department of Energy. This work used resources of the Argonne Leadership Computing Facility, which is a DOE Office of Science User Facility supported under Contract DE-AC02-06CH11357. An award of computer time was provided by the Innovative and Novel Computational Impact on Theory and Experiment (INCITE) program. This research is also part of the Blue Waters sustained-petascale computing project, which is supported by the National Science Foundation (awards OCI-0725070 and ACI-1238993) and the state of Illinois. Blue Waters is a joint effort of the University of Illinois at Urbana-Champaign and its National Center for Supercomputing Applications.

\bibliographystyle{apj}
\bibliography{main}

\end{CJK*}
\end{document}